\documentclass{moriond}

\usepackage{bold-extra}

\def\Journal#1#2#3#4{{#1} {\bf #2}, #3 (#4)}

\def\be{\begin{equation}}
\def\ee{\end{equation}}
\def\bea{\begin{eqnarray}}
\def\eea{\end{eqnarray}}

\newcommand{\Photo}{}

\begin{document}
\vspace*{4cm}
\title{CONSTRAINING MODELS WITH EXTRA HEAVY GAUGE BOSONS USING LHC MEASUREMENTS}

\author{ M.M.~Altakach}

\address{Laboratoire de Physique Subatomique et de Cosmologie,
 Universit\'e Grenoble-Alpes, CNRS/IN2P3,\\
 53 Avenue des Martyrs, 38026 Grenoble, France \\
 Institut f\"ur Theoretische Physik, Westf\"alische
 Wilhelms-Universit\"at M\"unster, Wilhelm-Klemm-Stra\ss{}e 9, 48149
 M\"unster, Germany}

\maketitle\abstracts{
We discuss significant improvements to our calculation of electroweak (EW) $t\bar{t}$ hadroproduction in extensions of the Standard Model (SM) with extra heavy neutral and charged spin-1 resonances using the \texttt{Recola2} package.
We allow for flavour-non-diagonal $Z'$ couplings and take into account non-resonant production in the SM and beyond including the contributions with t-channel $W$- and $W'$-bosons. 
We include next-to-leading order (NLO) QCD corrections and consistently match to parton showers with the POWHEG method fully taking into account the interference effects between SM and new physics amplitudes. 
We briefly describe the Contour method and give some information about the \texttt{Rivet} repository which catalogues particle-level measurements subsequently used by \texttt{Contur} to set limits on beyond the SM (BSM) theories.
We explain how we use our calculation within \texttt{Contour} in order to set limits  on models with additional heavy gauge bosons using LHC measurements, and illustrate this with an example using the leptophobic Topcolour (TC) model.
}

\section{NLO QCD corrections to EW $\mathbf{t\bar{t}}$ hadroproduction beyond the Standard Model}
\label{section-1}

The SM of particle physics is  based on the ad hoc $SU(3)_C \times SU(2)_L \times U(1)_Y$ gauge group.
The unification of this group in a larger, simple one, is theoretically very attractive. In this view, the SM is seen as an effective model valid at low energies. Thus, given the fact that the possible unification groups have a rank greater or equal than the SM, additional subgroups like a $U(1)$ or a second $SU(2)$ may appear at an intermediate stage when the unification group is broken down to the SM group. Interestingly, a new $U(1)$ group factor predicts one additional gauge boson generally denoted $Z'$ in the literature. On the other hand, since $SU(2)$ is a non-Abelian group with three generators, an additional one leads to three gauge bosons of which one is neutral and two are charged and denoted $W'^{\pm}$.

In many cases, the $Z'$ and $W'$ resonances decay leptonically and
the strongest constraints come from searches with dilepton final states. However, top-quark observables are still very interesting.
Indeed, the heavy top-quark may play a special with respect to EW symmetry breaking and to new physics that couples preferentially to the third generation, or does not couple to leptons at all. In fact, even if the resonances can couple to leptons, adding top-quark observables can be important to distinguish between different BSM scenarios.

In 2015, a calculation of NLO QCD corrections to the 
electroweak $t\bar{t}$ production in the presence of a diagonal $Z'$ resonance  \cite{Bonciani:2015hgv} was performed. 
Recently, we have redone the calculation including a number of major improvements \cite{altakach:1}. We allowed for flavour non-diagonal/generation non-universal couplings between the $Z'$-boson and the SM quarks. Thus, $Z'$-models explaining the $B$-flavour anomalies  can now be studied. Furthermore, we added the $t$-channel $W$ and $W'$  non-resonant contributions. As before,  the photon induced channels for the SM were included in order to treat the initial state QED singularities, and all interference terms between the SM and the new physics were taken into account.
The amplitudes have been calculated using the \texttt{Recola2} package which computes tree and one-loop amplitudes at NLO (EW, QCD) in the Standard Model and beyond. As in the previous calculation, the amplitudes were matched to parton shower (PS) at NLO+PS accuracy within the \texttt{POWHEG\,BOX} framework. This was the first use case of \texttt{Recola2} BSM amplitudes in  a NLO+PS matched calculation. 

\section{Testing new-physics models with global comparisons to collider measurements: the \texttt{Contur} toolkit}
\label{section-2}

Working at the interface of theory and experiment in particle physics demands a balance between what we wish to understand regarding the theory under consideration, and what we can achieve using the experimental data. Even though an important number of measurements at the LHC were designed to deal with SM processes, they can still implicitly involve information about possible contributions from BSM physics. The method, ``Constraints On New Theories Using Rivet'', Contur \cite{Buckley:2021neu}, uses the fact that unfolded particle level measurements created in fiducial regions of the phase space are highly model-independent. These measurements can thus be exploited to get information about BSM processes implemented in Monte Carlo generators in a very generic way.

\begin{figure}[h!]
\begin{minipage}{0.5\linewidth}
\centerline{\includegraphics[width=\linewidth]{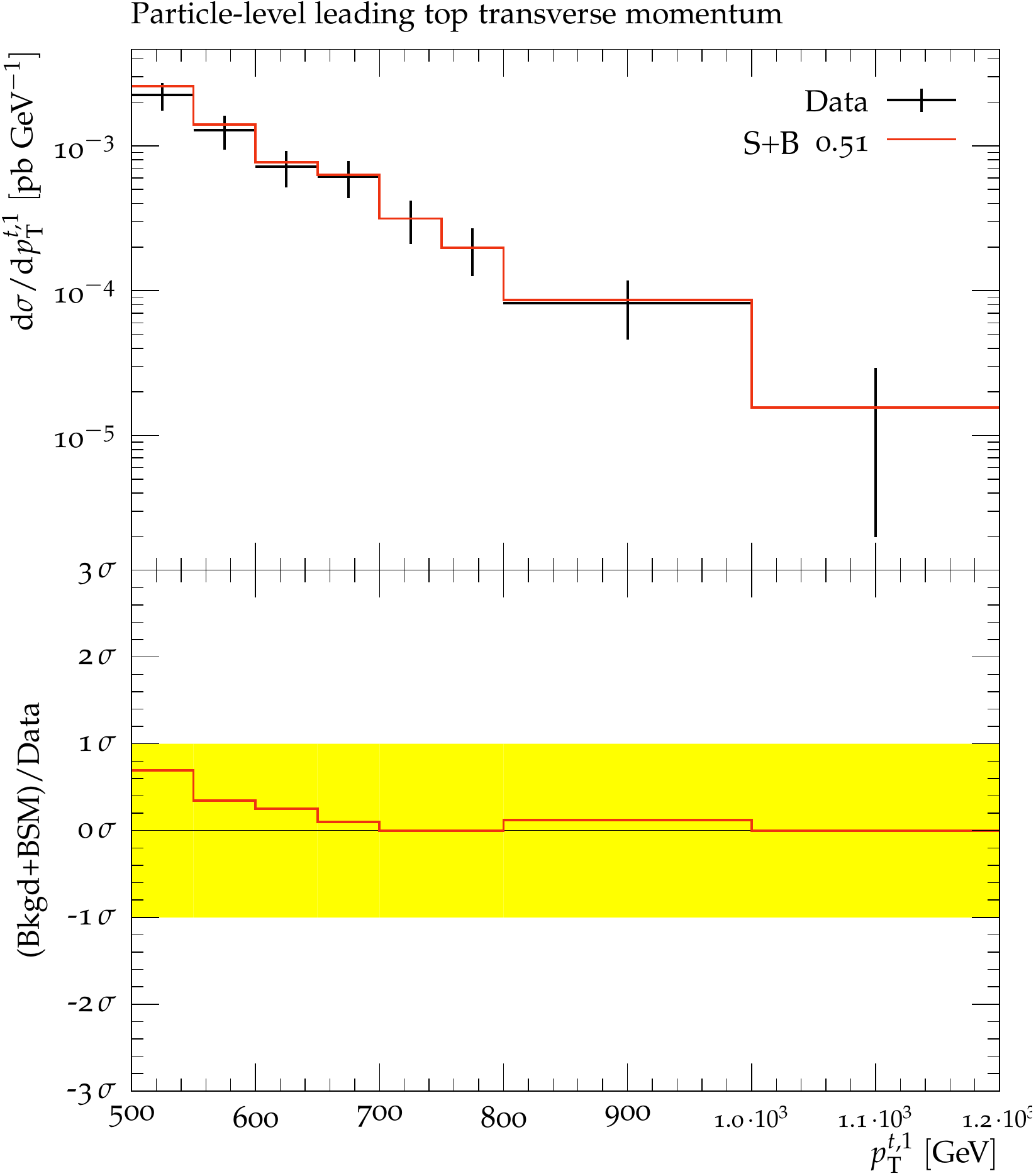}}
\end{minipage}
\hfill
\begin{minipage}{0.5\linewidth}
\centerline{\includegraphics[width=\linewidth]{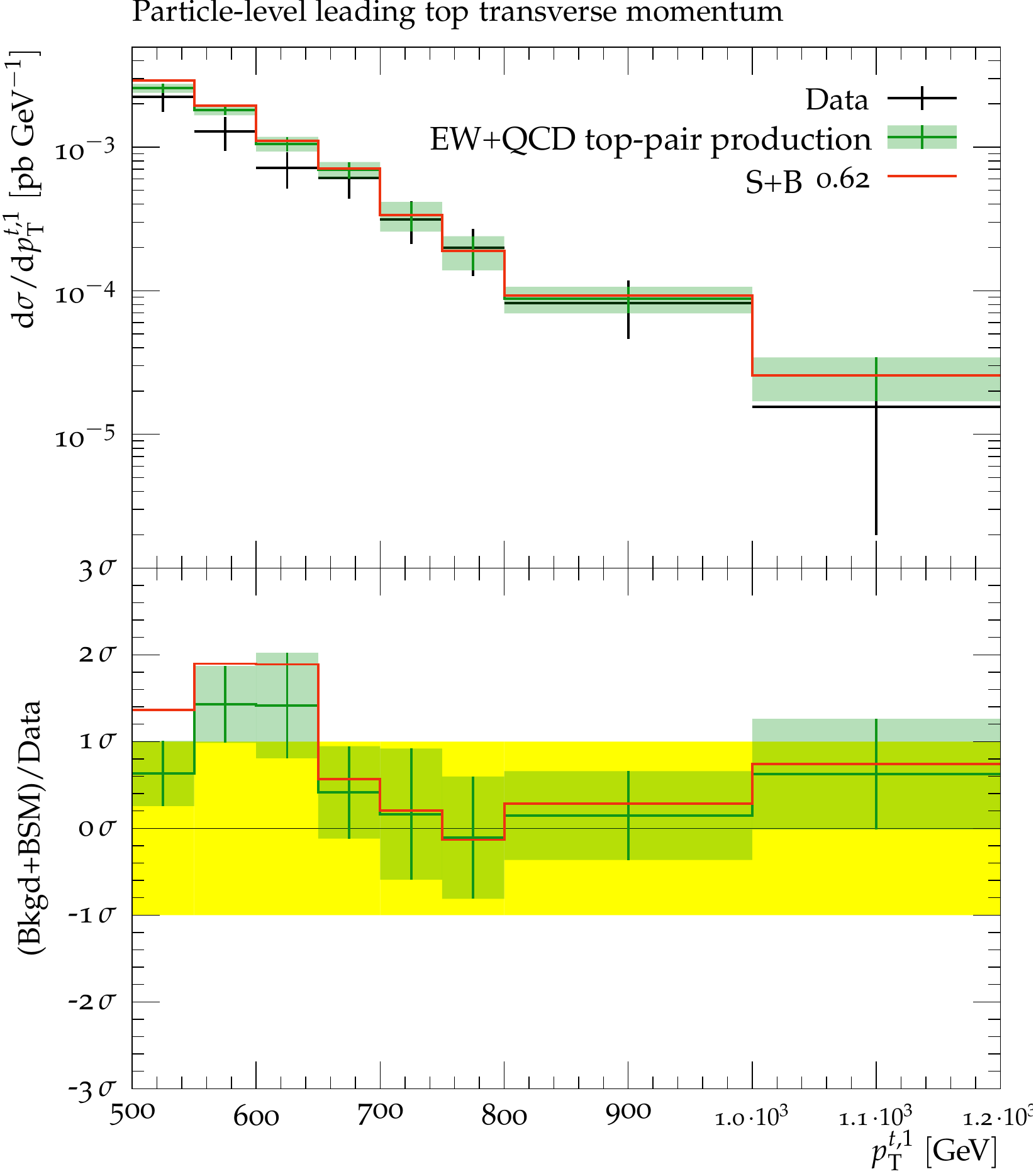}}
\end{minipage}
\hfill
\caption[]{A comparison of a leptophobic TC model \cite{lepTC} producing a $Z'$ signal. The black data points represent the observed data from the original measurement. An NLO QCD+EW $t\bar{t}$ background prediction is shown in green for comparison. The figure on the left shows the background model generated from data, with the one on the right showing the effect of using the theoretically calculated background model.}
\label{fig:Contur-method}
\end{figure}

Before proceeding with \texttt{Contur}, let us briefly introduce the tool that it uses. The \texttt{Rivet} \cite{rivet},
 or Robust Independent Validation of Experiment and Theory, repository is a system where particle-level measurements from the LHC and other colliders are preserved. Each analysis paper has its own \texttt{Rivet} analysis routine
which produces histograms that can be compared to the published plots in the corresponding paper. In other words, a \texttt{Rivet} routine chooses the generated events that would go into the fiducial region, and project their properties into histograms that have the same observables and binnings as in the measurement.

\texttt{Contur} exploits the set of \texttt{Rivet} analysis plugins to set the fiducial cuts on the BSM events instead of the SM ones. Two approaches can be used in order to calculate the exclusion limits with \texttt{Contur}: (i) The most complete one is when the background is defined by the simulated SM predictions and their associated uncertainties. In this case, the BSM signal is added on top of the background. The exclusion is obtained by comparing  the signal + background to the data within uncertainties.
(ii) The less complete approach is when the background is assumed to be the data. In this case, the BSM signal is superimposed on top of the data and the uncertainties on the data are considered to define the room that is left for the signal.
Both approaches are exemplified in figure~\ref{fig:Contur-method} for one point in the parameter space where on the left hand side (background = data) the signal (red) is 51\% excluded whereas on the right hand side (background = SM prediction (green)) the BSM signal (red) is excluded at 62\% CL.  \texttt{Contur} provides the book-keeping and steering machinery to repeat this process over a grid of parameter values.

\section{Constraints on theories with extra $\mathbf{Z’}$-boson in the $\mathbf{t\bar{t}}$ final state using \texttt{Contur}}

In order to set limits on models with an additional $Z'$-boson in the top-quark pair final state we use our calculation (see section~\ref{section-1}) within \texttt{Contur} as follows: the range for each input parameter of a chosen model can be specified using \texttt{Contur}.  Accordingly a \texttt{POWHEG} input file for every parameter point will be produced. Then, for each input file, \texttt{POWHEG} generates events in the Les Houches format. The events are then showered and transformed to the HepMC (needed by \texttt{Rivet}) format using \texttt{Pythia}. \texttt{Rivet} filters the HepMC events that would enter the fiducial region. Finally, following one of the two approaches described in section~\ref{section-2}, \texttt{Contur} compares the size of any deviation to the background for each set of parameters and  gives an exclusion limit. If desired \texttt{Contur} can as well combine the resulting limits of all the sets into one map.

To exemplify this chain we will use the leptophobic Topcolour model \cite{lepTC}. This model has two $SU(3)$ symmetries of which one couples to the first and second fermion generations whereas the other one only couples to the third generation. The two groups are then broken down to the $SU(3)_C$ of the SM. In order to prevent the bottom-quark from being as massive as the top-quark, the TC model involve an additional $U(1)_2$ symmetry, and as we mentioned in section~\ref{section-1}, this will lead to an additional $Z'$-boson. The $U(1)_1\times U(1)_2$ is broken down to the $U(1)_Y$ of the SM. The resulting $Z'$ is chosen to only couple to the first and third quark generations and, as can be understood from the name of the model, to not couple to leptons at all. Beside the mass of the new gauge boson, this model has three free parameters: the ratio of the two $U(1)$ coupling constants, $\cot \theta_H$, as well as the relative strengths
$f_1$ and $f_2$ of the couplings of the right-handed up- and down-type quarks with respect to those of the left-handed quarks. We set $f_1$ and $f_2$ to 1 and 0, respectively, and calculate $\cot \theta_H$ with respect to the total decay width of  the $Z'$-boson, which is given in Eq.~6 of Ref.~\cite{lepTC}.

Using the tool-chain that we described at the beginning of this section, we now show the resulting map for a specific range of parameters of the leptophobic TC model. This was done following the second approach of the Contur method (background = data). We consider a range of $Z'$ masses $m_{Z'} \in [1000, 5000]$ GeV and we calculate $\cot \theta_H$ such that $\Gamma_{Z'} / m_{Z'} = \{ 0.01, 0.039, 0.068, 0.097, 0.126, 0.155, 0.184, 0.213, 0.242, 0.271, 0.30 \}$. For the proton PDFs, we use the NLO luxQED set of NNPDF3.1 as implemented in the LHAPDF library (ID = 324900), and we choose equal values for the factorisation and renormalisation scales, $\mu_F$ and $\mu_R$ respectively, which we identify with the partonic centre-of-mass energy: $\mu_F = \mu_R = \sqrt{\hat{s}}$.
\begin{figure}[h!]
\begin{minipage}{\linewidth}
\centerline{\includegraphics[width=0.7\linewidth]{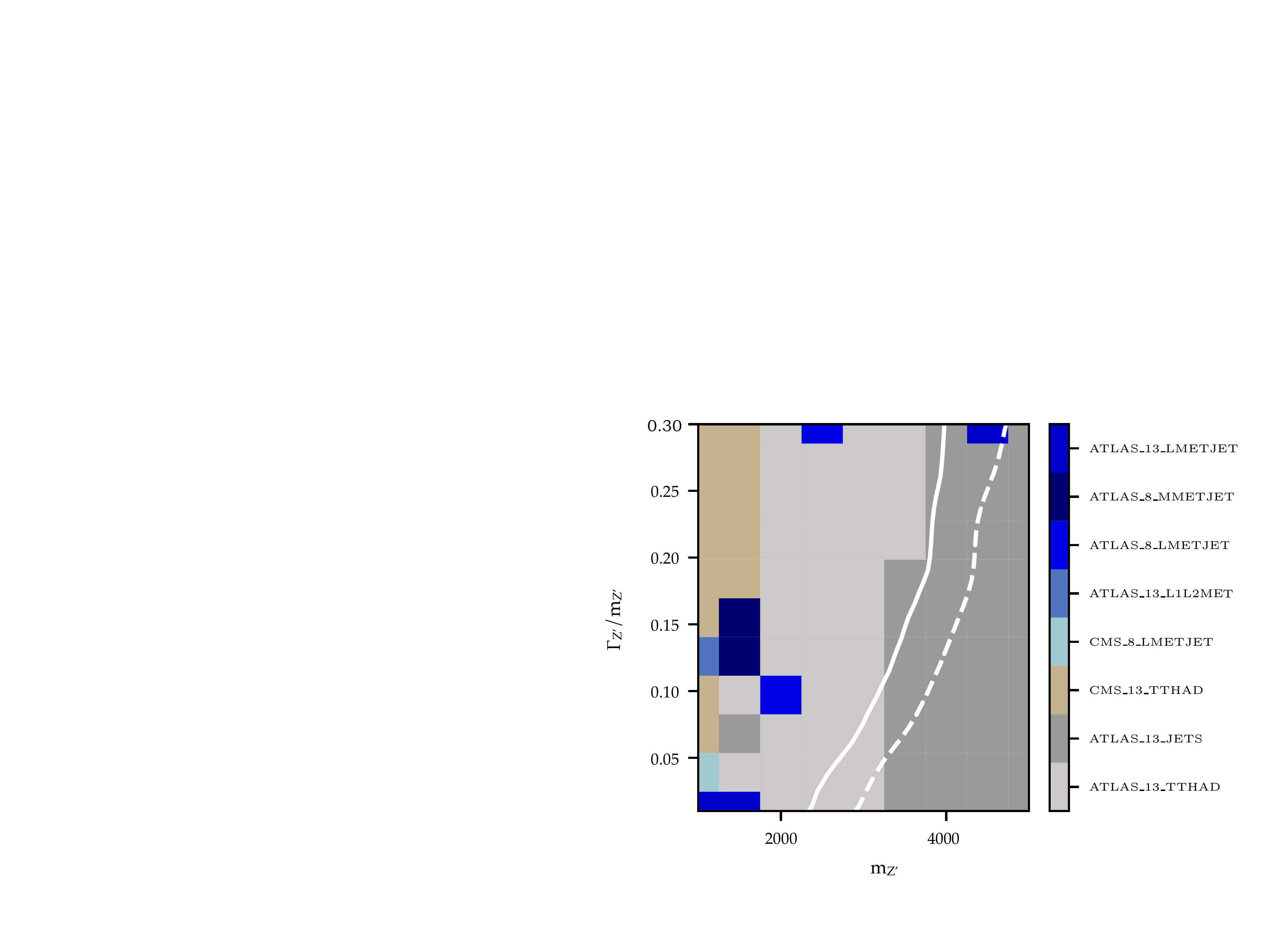}}
\end{minipage}
\hfill
\caption[]{A comparison of a leptophobic TC model \cite{lepTC} producing a $Z'$ signal. The black data points represent the observed data from the original measurement. An NLO QCD+EW $t\bar{t}$ background prediction is shown in green for comparison. The figure on the left shows the background model generated from data, with the one on the right showing the effect of using the theoretically calculated background model.}
\label{fig:TC-results}
\end{figure}
The resulting map for the combined exclusion limits can be seen in figure~\ref{fig:TC-results} where the solid white line represents the exclusions at $95\%$ CL and the dashed white line represents the $68\%$ CL exclusions. In order to prevent the risk of over-stating the sensitivity to a given signal, \texttt{Contur} divide the measurements of \texttt{Rivet} into orthogonal pools based on the experiment which performed the measurement, the centre-of-mass
energy of the LHC beam, and the final state which was probed. The results from each pool can then be combined without the risk of double counting. The colours in figure~\ref{fig:TC-results} indicate the dominant pool for each scan point. We see that the most sensitive analysis pools in the area between the solid and dashed lines are the $13$ TeV ATLAS all-hadronic boosted $t\bar{t}$ and inclusive jet and dijet cross section measurements.

\section*{Summary}

We discussed some aspects of our new calculation of NLO QCD corrections to EW top pair production using the \texttt{Recola2} package in the presence of $Z’$ and $W’$ bosons. 
We then briefly described the Contur method and the \texttt{Rivet} repository. Finally, we showed a complete chain of tools that allows us to obtain exclusion limits for BSM models with an extra $Z’$-boson in the top-quark pair channel, and exemplified it using the leptophobic TC model. We saw that the limits at $95\%$ and $68\%$ CL on the mass of the $Z'$-boson come mostly from the all-hadronic boosted $t\bar{t}$
and the inclusive jet and dijet cross section ATLAS measurements at $13$ TeV.

\section*{Acknowledgments}

I am grateful to J.M.~Butterworth, T.~Je\v{z}o, M.~Klasen, and I.~Schienbein for their ongoing collaboration on the work presented here. This work was supported by the BMBF under contract 05H18PMCC1,  the DAAD, the IN2P3 project “Théorie – BSMG”, and the European Union’s Horizon 2020 research and innovation programme as part of the Marie Sk lodowska-Curie Innovative Training Network MCnetITN3 (grant agreement no.~722104).

\end{document}